\documentclass[epj,nopacs]{svjour}
\usepackage{graphicx}
\usepackage[colorlinks,breaklinks]{hyperref}
\hypersetup{linkcolor=blue,citecolor=blue,filecolor=black,urlcolor=blue} 
\usepackage[numbers]{natbib}

\begin{document}
\hyphenation{off-line}
 \title{The ATLAS Data Quality Defect Database System}
 \author{T. Golling\inst{4} \and H.S. Hayward\inst{2} \and P.U.E.~Onyisi\inst{1}\mail{ponyisi@hep.uchicago.edu} \and H.J.~Stelzer\inst{3} \and P.~Waller\inst{2}
}
 \institute{Enrico Fermi Institute, University of Chicago, Chicago, IL 60637, U.S.A. 
\and Oliver Lodge Laboratory, University of Liverpool, Liverpool L69 3BX, U.K. 
\and Department of Physics and Astronomy, Michigan State University, East Lansing, MI 48824, U.S.A.
\and Department of Physics, Yale University, New Haven, CT 06520, U.S.A.}
\date{\today}
\abstract{The ATLAS experiment at the Large Hadron Collider has implemented a new system for recording information on detector status and data quality, and for transmitting this information to users performing physics analysis.  This system revolves around the concept of ``defects,'' which are well-defined, fine-grained, unambiguous occurrences affecting the quality of recorded data.  The motivation, implementation, and operation of this system is described.
}
\maketitle
\section{Introduction}
The ATLAS detector at the Large Hadron Collider (LHC) \cite{Aad:2008zzm} is a complex general purpose particle detector with approximately 100 million readout channels.  In common with many modern physics experiments it combines a large number of distinct subcomponents: it features nine major detection technologies and a number of special-purpose systems.  The data from specific components may not be usable for physics studies for certain periods of time.  For example, a component may be at a non-nominal voltage, readout electronics may need to be reset, or the data may be noisier than usual.  These situations arise both from the standard operation procedure and from unexpected failures.  Because not all physics studies rely on all components and these issues are often transient, it is desirable to continue data acquisition even in a degraded state.  It is also possible for data to be badly calibrated or otherwise not handled properly in the offline reconstruction, although possibly recoverable later using updated software or calibrations.  The ability to use more data by ignoring unnecessary components is not a trivial effect: of 1.25 fb$^{-1}$ of data recorded by ATLAS between March and June 2011 at a center of mass energy of 7 TeV, analyses used between 1.04 and 1.21 fb$^{-1}$ depending on which detector components were required.  

For physics analysis it is essential to know about these degraded conditions and to be able to exclude data from periods where detector problems would affect measurements.  Therefore the state of the detector (or the ``data quality'') must be monitored, recorded, and propagated to analysts.  This task involves both core data management issues and human interface concerns.  The detection of many problems is not fully automated and manual input is required.  The opportunity for incorrect data entry or wrong interpretation must be minimized.  The final decisions about what data to reject are often made long after the data are recorded, once the impact of various problems is better understood, so maximum flexibility should be a goal.  Analysts should be able to access the current best assessment of what data to use easily, while still being able to perform detailed queries on detector status when necessary.

A ``flag''-based data quality assessment chain implementation \cite{Adelman:2010zza}, similar in concept to those used in previous and current experiments (for example CMS \cite{Tuura:2010zza}), was in place at the start of ATLAS physics data collection.  The main information stored in this system was \textit{decisions} about whether the data recorded at a given time was usable for analysis.  This framework was used to produce the physics results of the 2010 data period.  However it became apparent that this flag system was inflexible and hard to handle in practice.  We therefore replaced this system during the winter 2010-2011 LHC shutdown with a new one where the stored information is the \textit{problems} that might go into making a decision, with the decisions on whether to use the data or not moved to overlying (stored) logic.  This seemingly simple change has made the evaluation of data quality at ATLAS much smoother; by tracking issues at a lower level than before, the overall process has been simplified.  In this paper we describe the features of the new ``defect''-based system and the improvements made over the flag system.

\section{The Data Quality Assessment Infrastructure and Process}
In this section we describe aspects of ATLAS experimental operation relevant to data quality monitoring, the basic database framework used for storing data quality information, and the final output of the data quality evaluation process.

The fundamental time granularity unit of detector configuration and status accounting in ATLAS is the ``luminosity block'' (LB).  These are sequential periods within a run assigned by the trigger hardware and embedded in the data stream for each recorded collision.  Their length is flexible (typically one minute long for 2011 data) and certain actions, such as a trigger configuration change request, will cause the start of a new luminosity block.

Time-dependent configuration, status, and calibration (``conditions'') information for ATLAS is stored in Oracle and SQLite databases using the COOL technology developed by the LCG project \cite{COOL,Verducci:2008zzb}.  A COOL ``folder'' consists of a set of ``channels'' sharing a folder-specific ``payload'' data structure, adapted to the information being stored (such as voltages, beam position, trigger configuration, and so on).  Channels have a numeric ID, name, and description associated with them.  Payloads can be stored on a channel-by-channel basis for specified ``intervals of validity'' (IOVs).  The start and end of an IOV are 63-bit integers, which in ATLAS are used to encode (run, LB) pairs or timestamps.  The information stored in COOL databases may be versioned via the ``tag'' mechanism: each tag acts as an independent set of IOVs and payloads for the channels of a folder.  Tags can be ``locked'' to prevent their data from being altered and guarantee reproducibility.  Data quality information is entered first in the special \texttt{HEAD} tag before being copied to other tags.

A typical ATLAS run \cite{Onyisi:2010jm} begins before protons are injected into the LHC and ends after the beams have been removed from the machine.  Outside of the ``stable beam'' period, when it is considered safe to run sensitive detectors in data-taking mode, the sensitive detectors are operated in a standby mode with reduced voltages and different readout configurations.

During data taking, a number of online applications record the status of the ATLAS detector in the conditions database, including the trigger and data acquisition system (TDAQ) \cite{L1,TDAQ}, the detector control system (DCS) \cite{BarriusoPoy:2008zz}, and the online data quality monitoring framework (DQMF) \cite{Kolos:2008zz,CuencaAlmenar:2011zz}.  The events from a specific set of triggers that are useful for detector monitoring are fed into an ``express stream'' which is promptly reconstructed in the ATLAS Tier-0 farm \cite{Elsing:2010zz}.  As part of the reconstruction, monitoring plots are produced and distributed, and automated checks are performed on these plots by the offline DQMF \cite{Adelman:2010zza}.  Various detector experts and physicist ``shifters'' review the information available to them and provide data quality feedback.  They also use information from the monitoring to improve the calibrations used for the reconstruction of events from all triggers that starts 36 hours after the end of a run.

Runs sharing similar conditions are grouped into ATLAS run periods and subperiods.  Subperiods may be as short as one run, if for example there is a rapid evolution of the LHC beam structure between runs.  After a subperiod is closed, it is given an additional review by detector experts, who sign off on the data quality assessment, certifying that all the runs have been inspected and all problems identified.  At this point the data are released for analysis.  A similar process is used after a reprocessing of previously-taken data with updated software.

The main end product of the ATLAS data quality infrastructure is a set of ``good run list'' (GRL) files which contain the list of luminosity blocks approved for analysis.  Several GRLs are produced, with different subdetectors required to be good depending on the needs of the corresponding physics studies.  These are the final products of the data quality assessment process that are delivered to users, who use the file recommended for their class of analysis.  The files use a common ATLAS XML interchange format, which is also used for example by the file provenance metadata architecture and the event-level metadata database \cite{Gallas:2010zz}.

\section{Data Quality Databases in 2010 Operation \label{sec:2010}}
The data quality databases implemented for 2010 operation \cite{Waller:2010zz} used a flag concept, where several different flag colors were used to reflect detector subcomponent status: green (ok), yellow (caution), red (bad), black (disabled), and grey (undecided).  There were $\mathcal{O}(100)$ components to be flagged for every run.  As the flags corresponded to specific subcomponents, the list of flags had very few changes after its initial definition.  Several COOL folders were used, each containing flags from different sources (online and offline DQMF monitoring, DCS monitoring \cite{Aad:2010zz}, online and offline physicist shifters).  Information from the different folders was merged to form the final output, which was primarily based on the flags set by the offline physicist experts and shifters.  Flags to be used for analysis were copied to dedicated COOL tags.

Several chronic issues were encountered with this system in operation:
\begin{enumerate}
 \item The set of problems that corresponded to each flag and color was not self-documenting.  Analysis users were largely unaware of what conditions caused data to be included and excluded from the GRLs and this information was not easy to discover.  As multiple problems could result in the same flag color, a lot of training was necessary to ensure that different shifters and experts applied uniform criteria; inevitable personnel change thus posed a long-term consistency concern.
 \item All issues needed to be reduced within days to a limited and unchanging set of possible flag and color combinations.  This required  immediate judgment of the likely impact of newly-found problems on physics analysis.  Several times, further investigation revealed the initial decisions to be incorrect, requiring retroactive changes to the database.
 \item Only storing the flag colors meant that a lot of useful information was not preserved.  Without resorting to looking at more basic sources (e.g.\ monitoring histograms), detailed information was at best provided in the free-form text comment field of the flag payload.  The only way to try to obtain lists of LBs subject to specific issues was to perform a text search, with attendant complications.
 \item The yellow flag proved troublesome.  Instead of only having to define the single green/red boundary, we instead had to define both green/yellow and yellow/red boundaries.  In fact, for the COOL tags used to generate analysis GRLs, yellow flags were not permitted, in order to reduce confusion.  All yellow flags were required to be ``resolved'' to green or red.  The semantics of yellow in the \texttt{HEAD} tag shifted over time from ``caution'' to ``expected recoverable''.  As a result the relationship between the flags in the \texttt{HEAD} and analysis COOL tags was often not obvious.
 \item There was no single authoritative list of data quality flags.  Lists were hard coded in several locations and adding a channel required a new ATLAS software release (and caused forward compatibility problems with older releases).
\end{enumerate}

It was decided to develop and implement an alternative system to address these difficulties.

\begin{figure}
\includegraphics[width=\linewidth]{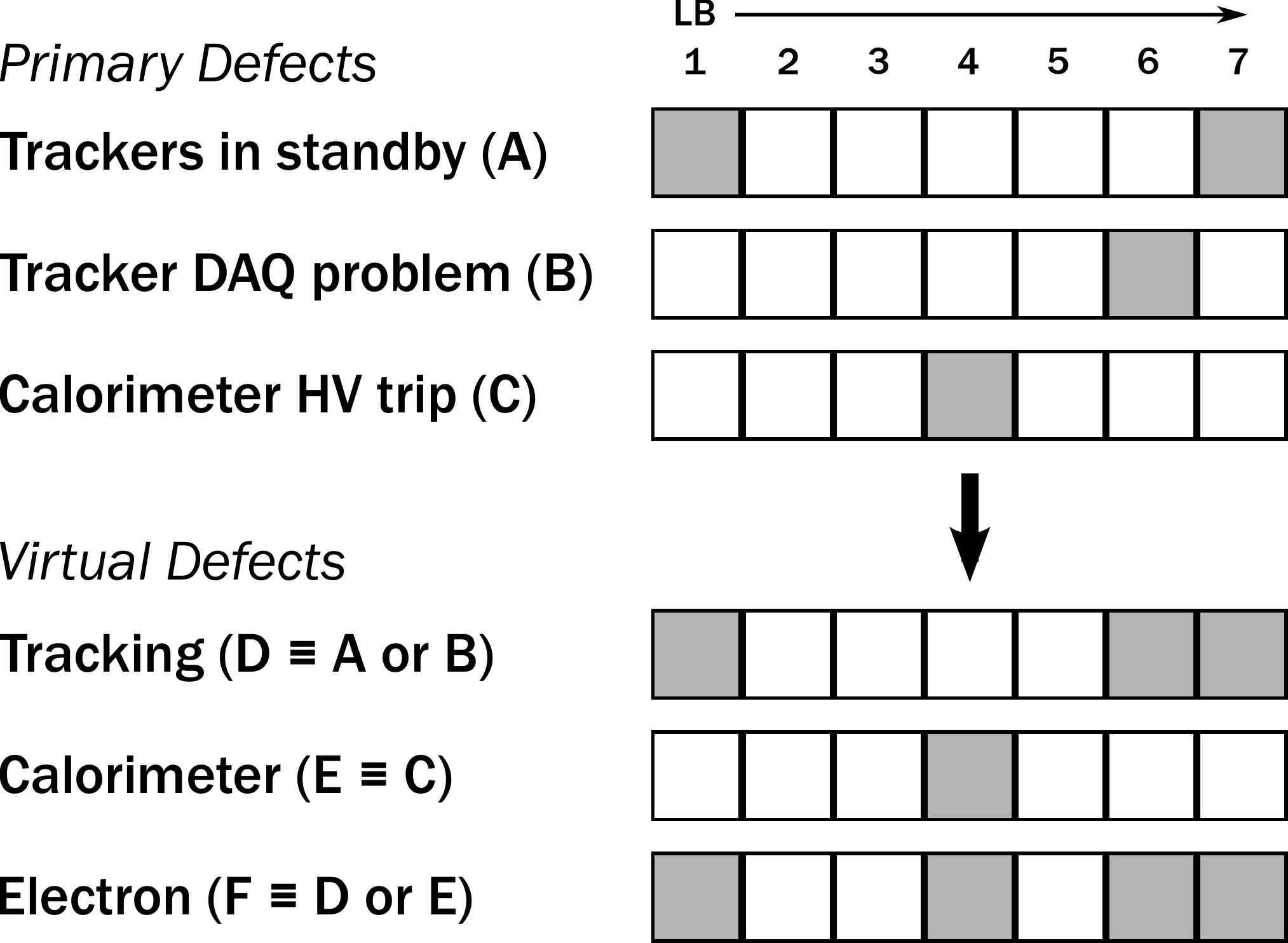}
 \caption{\label{fig:virtualdefect}A demonstration of how information is propagated from primary to virtual defects.  A simplified set of defects is shown, along with their states for various luminosity blocks during a run.  Shaded boxes indicate luminosity blocks in which the primary or virtual defect is reported to be present and corresponding events are to be rejected.  An analysis would depend only on the Electron virtual defect, only referring to ``deeper'' defects if it had unusual requirements.}
\end{figure}

\begin{figure*}
\includegraphics[height=3.9cm]{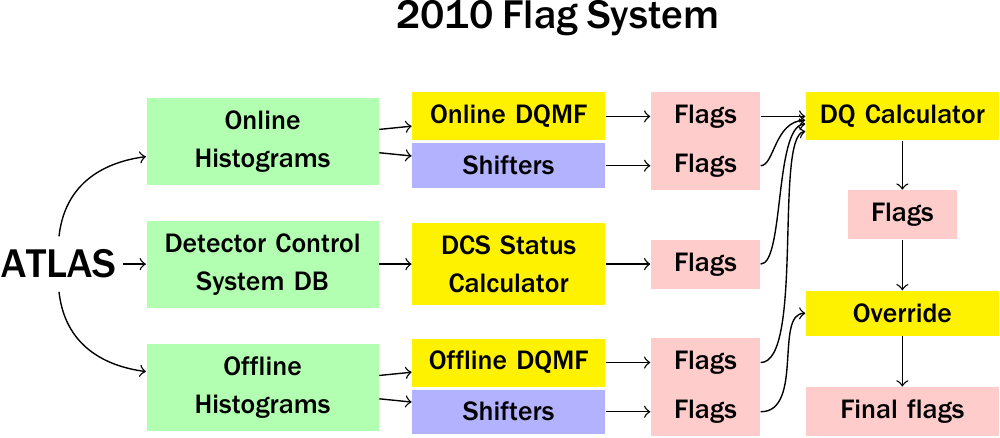}\hfill%
\includegraphics[height=3.9cm]{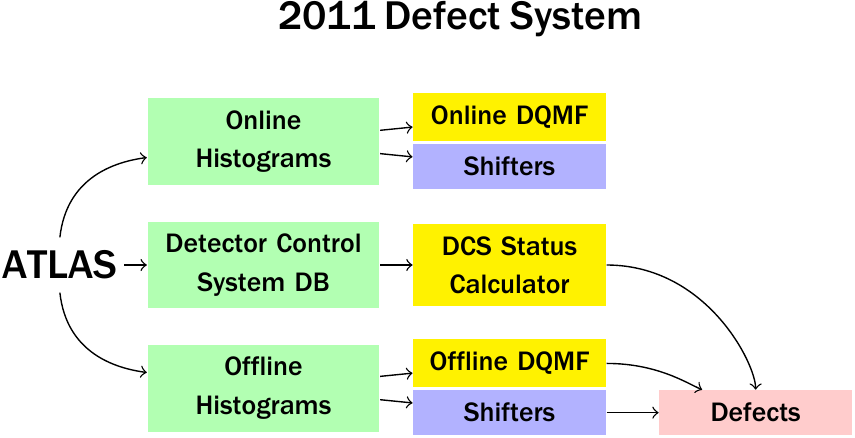}
\caption{\label{fig:comparison}A comparison of the information flow from data taking to physics analysis for the flag system used in 2010 data (left) and the defect system of 2011 data (right).  The final output used for constructing good run lists is in the bottom right in both cases.  The defect system is less complex than the flag system.  Some flags are still present in 2011 operation to ease the transition, but their use is deprecated.}
\end{figure*}

\section{Concepts of the Defect Database}
A ``defect'' is a deviation from a nominal detector condition.  A defect is either present or absent for a given luminosity block.  An arbitrary number of defects may be defined.  

A defect may be explicitly stored in a database or be computed on retrieval.  Defects whose values are stored in the database are referred to as ``primary defects'' to distinguish them from ``virtual defects,'' which are defined combinations of primary defects or other virtual defects and only computed on access.  Primary defects are those that are input to the system on a day-to-day basis, while virtual defect definitions evolve much more slowly.  

A virtual defect is specified by the other defects (primary or virtual) that it depends on.  If any of its dependencies are present, a virtual defect is present for a luminosity block (the presence of primary and virtual defects has the same semantics).  Virtual defects are used to combine primary defects into higher level concepts; for example, all muon trigger defects that are serious enough to exclude data from use are combined in a single virtual defect.  The main purpose of virtual defects is to simplify defect database queries and to encapsulate the current best understanding of which primary defects correspond to problems where the corresponding data should not be used in physics analyses.  A demonstration of virtual defect logic is shown in Figure~\ref{fig:virtualdefect}.  A similar ``virtual flag'' concept existed for the flag system, but the combination logic was more complicated as flags had more possible states.

The values of the primary defects and the definitions of the virtual defects are stored and versioned with the COOL tag mechanism.  This ensures the reproducibility of database queries, while allowing defect values and virtual defect definitions to evolve as necessary.  Within a single COOL tag, a virtual defect has a constant definition for all runs.  The virtual defect definitions can be updated independently of the primary defect information as the understanding of the effect of detector problems improves.  Because of this both the relevant primary and virtual defect tags must be specified during a retrieval.  

The flag system had a number of different parallel COOL folders storing information from different sources, which were merged to determine the final flags.  We considered this unnecessary for the defect database, as any given defect should either be reliably automatically detected, or require manual input.  There is therefore only one production instance of the defect database, filled both by people and software, and no merging steps are required.

We emphasize that a defect need not be so serious as to cause data not to be used in analysis; it may serve as an issue tracking mechanism, or be mainly of interest for checks of possible systematic effects.  It is also possible to ignore specific primary defects during the virtual defect computation, again to facilitate studies of systematic uncertainties.

The defects carry some metadata with every entry, including a comment, the username of the person or ID of the automated process that filled the entry, and whether the problem is likely to be recovered later.

The defect database concept addresses the concerns of Section~\ref{sec:2010} as follows:
\begin{enumerate}
 \item There is one defect for each class of problem.  The meaning of the defect is explained in the description field of the defect; if this is done clearly enough there should be no ambiguity.
 \item A new type of problem immediately gets a new defect.  Its effect on the GRLs is handled by the virtual defects, which can be updated when a fuller picture of the impact of the problem is obtained.  It is also not necessary to anticipate all problems in advance, as defects can be added as problems occur.
 \item All the information that was used to make decisions with the flag system is now explicitly available and easy to query. In particular, it is simple to determine the set of all data in which a defect was present.
 \item The stored information is binary (a defect is either present or absent).  The ``expected recoverable'' meaning of the yellow flag is provided by a Boolean field in the defect.  As there is no longer a resolution process required, making a COOL tag of the defects to be used to generate good run lists is as simple as copying the \texttt{HEAD} information.
 \item The defect database is self-describing.  It was an explicit design requirement that the access application programming interface (API) should not add additional information beyond that in the database.
\end{enumerate}

\section{Implementation of the Defect Database}
The defect database is implemented with two COOL folders, one for the primary defect data and the other for the virtual defect definitions.  These two folders are versioned independently but their COOL tags can be tied together with the ``hierarchical tag'' mechanism, meaning only a single tag needs to be presented to the analysis users.

As an optimization to cope with the large number of expected defect channels, the absence of any data for a defect for an interval of validity is considered equivalent to an absent defect.  This optimization means that not only is the database smaller, but the demands on the shifters are reduced as well since they do not have to explicitly mark good data.

A single API, written in Python, has been created that covers the vast majority of defect database creation, filling, query, and manipulation needs. The Python library is implemented in 1.3 thousand lines of code (kloc).  An extensive suite of tests using the \texttt{nose} package \cite{nose} is run nightly to ensure that the library conforms to specifications.  As the specifications were clearly defined before the package was written, a test-driven process allowed rapid development over a few days with confidence in code correctness.  The API enforces certain validity conditions for input (e.g.\ virtual defects should only reference existing primary and virtual defects) and is the only approved input method for the defect database.  For use in event reconstruction, the standard ATLAS Athena \cite{Calafiura:2005zz} C++ interface library is used to directly access the database.

As the user interface software needed to be rewritten to handle the new defect system, we decided to take advantage of new Web 2.0 technologies to provide a more intuitive and responsive web application than the one previously used for the flag database.  The new shifter application consists of 0.4 kloc of backend Python code running in a CherryPy web application server and 1 kloc of client-side Javascript using the Google Closure framework, replacing the 5.3 kloc of PHP code comprising the old application.

The fact that the defect database is the authoritative source of all information concerning defects allows the creation of a single administrative web interface for defect management.  This interface allows defect creation, virtual defect creation and definition editing, and tag creation and updating.  This application, hosted in the same server process as the shifter application, consists of 0.4 kloc of backend Python code and 0.8 kloc of client-side Javascript.  There was no similar interface for the flag system.

\sloppy Several defects not corresponding to detector problems have been added for bookkeeping purposes.  A \texttt{NOTCONSIDERED} defect was initially set present for all luminosity blocks, and is then set absent for the LBs comprising a run when that run is reviewed by the data quality group.  Due to the convention that the absence of defects indicates that there is no problem, a guard defect like this is necessary to avoid including runs in GRLs that are not yet reviewed.  In addition, a set of \texttt{UNCHECKED} defects were created that serve as workflow management markers.  These defects are all automatically set present when a data-taking run completes, and are unset by the shifter signoff procedure.  Virtual defects that depend on the \texttt{UNCHECKED} defects will therefore reject data until the shifters and experts have reviewed it.  The administrative interface will not permit the generation of official good run lists for a run period if any \texttt{UNCHECKED} defects are present.

\fussy When transitioning from the flag system, we wanted to ensure minimal disruption to downstream consumers of data quality information.  The interface between the data quality database and the users lies primarily in the GRL generation mechanism.  We created new virtual defects with the same names as the old flags and grouped the new primary defects under these virtual defects.  The non-green flags from 2010 data were also imported as defects.  (A full retroactive filling of 2011 defects for 2010 was considered impractical.)  We were largely able to avoid changes to the GRL generation configurations and retain the ability to generate GRLs for 2010 data with the defect database.

A comparison of the information flow in the flag and defect database systems is shown in Figure~\ref{fig:comparison}.  Some of the flag system COOL folders are still being filled, but now have no direct impact on GRL creation.  As more confidence is gained with automatic detection of various problems, the relevant information is written directly into the defect database as well (implemented so far for portions of the DCS and offline DQMF information).

\section{Operation of the Defect Database}
The defect database has been used for the 2011 running.  Integration into the data quality assessment workflow was smooth and user feedback very positive.  As anticipated, new detector problems are entered into the database immediately, allowing their physics impact to be studied at a more relaxed pace while maintaining clear documentation of the affected data.  Anecdotal evidence suggests that the frequency of user input errors has been reduced substantially, and that the removal of the resolution phase when preparing COOL tags for analysis has reduced turnaround time allowing data analysis to begin sooner.  Care must be taken to avoid creating duplicate defects; this is achieved by restricting defect creation to a small set of experts.

\begin{figure}
\includegraphics[width=\linewidth]{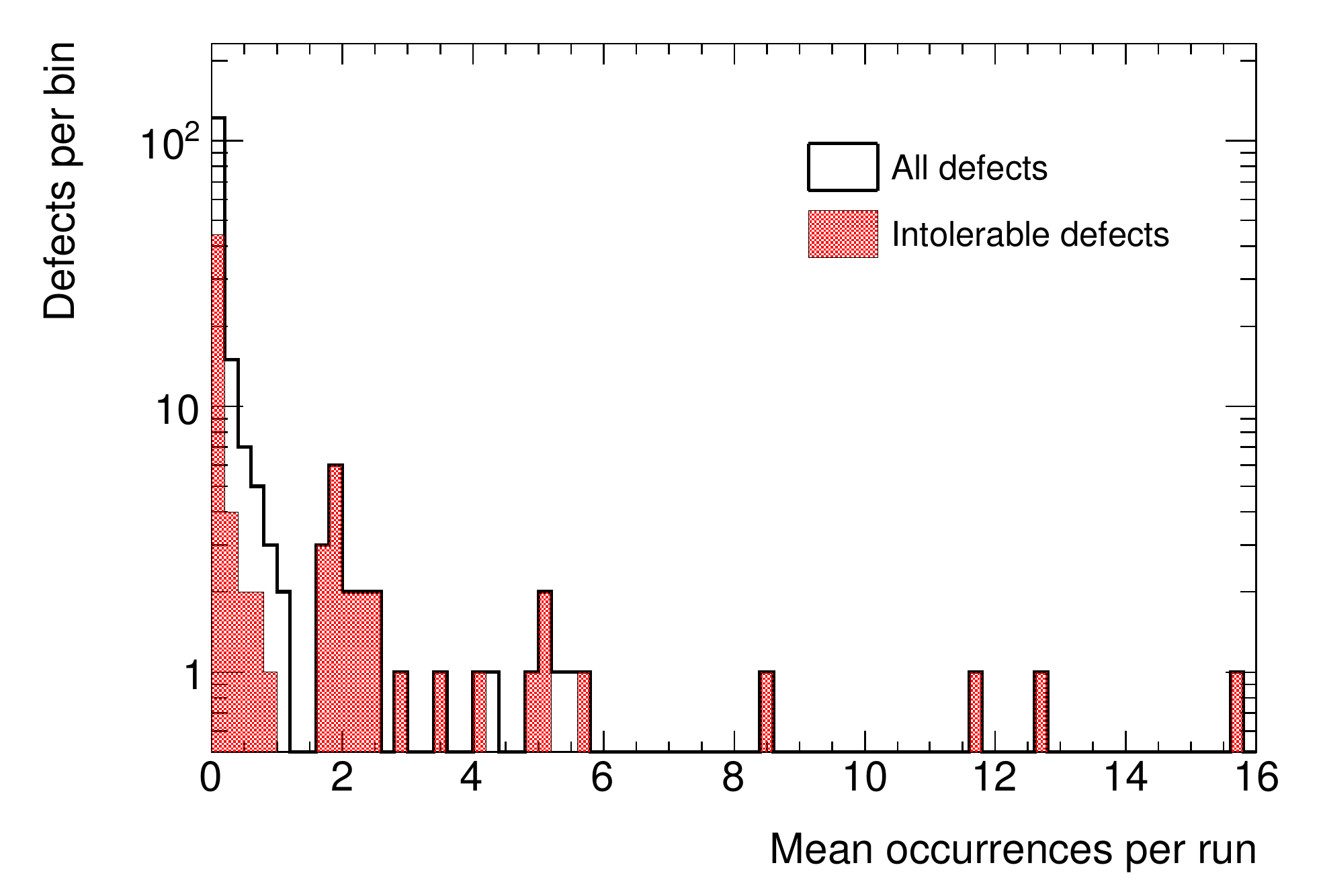}
\caption{\label{fig:iovs}A histogram of the mean number of occurrences (IOVs) recorded for each defect in runs available for physics analysis at 7 TeV center of mass energy between March and June 2011.  The peak near 2 occurrences per run is due to detector components being in standby at the start and end of runs.  ``Intolerable'' defects are those which will cause at least one analysis to reject the affected data.}
\end{figure}

As of the accumulation of 1.25 fb$^{-1}$ of data in June 2011, there were 619 defects and 172 virtual defects defined.  Including all COOL tags, the database contains approximately 33 MB of data, which promises good scalability for the future.
Figure~\ref{fig:iovs} shows the mean number of intervals of validity per run (of whatever length) defined for primary defects in runs available for physics analysis at 7 TeV center of mass energy between March and June 2011; this corresponds to the number of rows that are inserted into the database.  Most defects are rare and occur much less often than once per run.  The defects reflecting when various components are in a standby state create the peak at 2 IOVs per run.  There are a few defects that occur quite often, which reflect frequent but short (i.e.\ single LB) detector problems.

Querying the database is quite fast.  For example, querying all defects and virtual defects for the 1.25 fb$^{-1}$ of data recorded through June 2011 using the Python API takes less than 40 seconds, including the virtual defect computation.  A single virtual defect, such as the barrel electron quality, takes under five seconds.  To retrieve the full set of primary defects takes under a second, including database connection setup time.

\section{Conclusion}
The ATLAS experiment requires stringent documentation and tracking of detector problems that affect the usability of data for analysis.  We have implemented a ``defect database'' system that allows straightforward entry and retrieval of specific types of problems, as well as combinatoric logic to determine which data should not be used for analysis due to specified issues.  We have demonstrated that such relatively low-level issue tracking is practical even for an experiment of the complexity of ATLAS, and in fact more successful than storing only coarse decisions on the usability of data.

\begin{acknowledgement}
We thank our colleagues in ATLAS for their suggestions, encouragement, and cooperation during the construction of the defect system.  This work was supported by the U.S.\ National Science Foundation and the U.K.\ Science and Technology Facilities Council.  P.U.E.O.\ was partly supported by a Fermi Fellowship from the University of Chicago.
\end{acknowledgement}

\bibliographystyle{apsrev}
\bibliography{paper}

\end{document}